\documentclass[pra,aps]{revtex4}
\usepackage{graphicx}
\usepackage{colordvi}
\begin{document}
\title{Dynamics of imbalanced quasi-one-dimensional binary Bose-Einstein condensate in external potentials}
\author{K. K. Ismailov$^{1,3}$, B. B. Baizakov$^1$, F. Kh. Abdullaev$^{1,2}$ and M. Salerno$^3$}
\affiliation{$^1$ Physical-Technical Institute, Uzbek Academy of Sciences, 100084, Tashkent, Uzbekistan \\
$^2$ Department of Theoretical Physics, National University of
Uzbekistan, 100174, Tashkent, Uzbekistan \\
$^3$ Dipartimento di Fisica ``E.R. Caianiello", and INFN Gruppo
Collegato di Salerno, \\ Universit\'a di Salerno, Via Giovanni Paolo
II, 84084 Fisciano, Salerno, Italy}
\date{\today}

\begin{abstract}
In the framework of coupled 1D Gross-Pitaevskii equations, we
explore the dynamics of a binary Bose-Einstein condensate where the
intra-component interaction is repulsive, while the inter-component
one is attractive. The existence regimes of stable self-trapped
localized states in the form of symbiotic solitons have been
analyzed. Imbalanced mixtures, where the number of atoms in one
component exceeds the number of atoms in the other component, are
considered in parabolic potential and box-like trap. When all the
intra-species and inter-species interactions are repulsive, we
numerically find a new type of symbiotic solitons resembling
dark-bright solitons. A variational approach has been developed
which allows us to find the stationary state of the system and
frequency of small amplitude dynamics near the equilibrium. It is
shown that the strength of inter-component coupling can be retrieved
from the frequency of the localized state's vibrations.
\end{abstract}
\pacs{67.85.Hj, 03.75.Kk, 03.75.Lm} \maketitle

\section{Introduction}

Two-component Bose-Einstein condensates (BEC) may show a variety of
interesting phenomena depending on the character and strength of
intra-species and inter-species forces \cite{kevrekidis2016}. Among
all possible settings the case of two self-repulsive BECs forming a
bound state due to inter-species attraction has been of special
attention. In two- and three-dimensional free space such a bound
state of binary gas is unstable against collapse or spreading
\cite{malomed-book}, though the stability provided by a
two-dimensional external periodic potential was reported
\cite{ma2016}. By contrast, in a one-dimensional free space stable
localized state can emerge, which is also known as a {\it symbiotic
soliton} \cite{adhikari2005,perezgarcia2005,adhikari2008}. The
peculiarity of the symbiotic soliton is that its individual
components cannot exist separately from each other, therefore its
dissociation into individual solitons cannot occur. This fact lies
in the core of their enhanced robustness and unusual dynamical
behavior during interactions with external potentials
\cite{javed2022} and mutual collisions. Although the properties of
symbiotic solitons have been reported in many publications, they
were mainly focused on balanced settings with all atoms bound in the
symbiotic soliton. The influence of excess atoms in imbalanced
mixtures upon the static and dynamic properties of symbiotic
localized states remains less explored. It should be noted, that
previous work on self-trapped localized states in imbalanced
ultra-cold atomic mixtures mainly was concerned with quantum
droplets \cite{tengstrand2022,flynn2022}, an issue that is beyond
the scope of this paper.

Our objective in this work is to show that stable localized states
similar to symbiotic solitons can exist in imbalanced binary
self-repulsive BECs when the number of atoms in one component
exceeds the number of atoms in the other component. We address the
situation when inter-component attraction slightly overcomes
intra-component repulsion so that the overall mean-field
nonlinearity is weakly attractive. A similar setting in a
three-dimensional case was employed to produce quantum droplets (see
review articles \cite{luo2021,guo2021,bottcher2021}). Dynamics of
symbiotic localized states will be considered in the harmonic trap,
rectangular potential well and toroidal confinement for quantum
gases.

This paper is organized as follows. In the next Section II, the
system of governing equations is introduced and a variational
approach for the binary BEC is developed. In Sec. III we consider
the dynamics of symbiotic localized states in imbalanced BEC
mixtures, confined to a parabolic potential, box-like and toroidal
traps. Sec. IV summarizes the main findings of this work.

\section{The model and variational approach}

The model is based on the following system of coupled
quasi-one-dimensional Gross-Pitaevskii equations (GPE)
\begin{eqnarray}
iu_t+\frac{1}{2}u_{xx}-V_1(x)u+\left(g_{11}|u|^2+g_{12}|v|^2\right)u &=& 0, \label{u} \\
iv_t+\frac{1}{2}v_{xx}-V_2(x)v+\left(g_{22}|v|^2+g_{12}|u|^2\right)v
&=& 0, \label{v}
\end{eqnarray}
where $u(x,t), v(x,t)$ are the mean-field wave functions of the
binary condensate, $V_{1,2}$ are external trapping potentials for
the components, $g_{11}, g_{22}$ are intra- and $g_{12}$
inter-component coupling coefficients. The dimensionless quantities,
entering these equations are scaled using the frequency of the
radial confinement $\omega_{\bot}$, atomic mass $m$ and radial
harmonic oscillator length $l_{\bot} = \sqrt{\hbar/m\omega_{\bot}}$
as follows: time $t \rightarrow t \omega_{\bot}$, space $x
\rightarrow x/l_{\bot}$, wave function $\{u,v\} \rightarrow
\sqrt{2|a_s|}\, \{u,v\}$. When considering imbalanced settings with
different number of atoms $N_{1,2}$ in the components, we shall
denote $u(x,t)$ and $v(x,t)$ as minority and majority components,
respectively.

In absence of external potentials, $V_{1,2}=0$, equal coefficients,
$g=g_{11}=g_{22}$, and equal number of atoms, $N=N_1=N_2$,
Eqs.(\ref{u})-(\ref{v}) admit an exact quiescent solution
\cite{ueda1990}
\begin{eqnarray}\label{exact}
u(x,t)=v(x,t)=A\,{\rm sech}\left[A\,\sqrt{g+g_{12}}\,x \right]
\exp\left(i\Theta \right), \quad \Theta=(A^2/2)(g+g_{12})\, t,
\end{eqnarray}
whose amplitude is linked to the norm $N=\int |u(x)|^2 dx$ as
$A=(N/2) \sqrt{g+g_{12}}$. The overall coupling constant is assumed
to be positive $g+g_{12} > 0$.

In presence of external potentials, analytic solutions are
unavailable and one has to recourse to approximate methods. For
general settings stationary solutions to Eqs.(\ref{u})-(\ref{v}) are
found in \cite{adhikari2008} by means of the variational approach
using Gaussian trial functions
\begin{equation} \label{ansatz}
u(x,t)=A_1\exp \left[ -\frac{x^2}{2a_1^2} + i b_1 x^2 + i\phi_1
\right], \quad v(x,t)=A_2\exp \left[ -\frac{x^2}{2a_2^2} + i b_2 x^2
+ i\phi_2 \right],
\end{equation}
where time dependent variational parameters $A(t), a(t), b(t),
\phi(t)$ define the amplitude, width, chirp and phase of the
components, respectively. Standard transformations of the
variational method \cite{anderson1983,malomed2002} with Lagrangian
density corresponding to Eqs.(\ref{u})-(\ref{v})
\begin{equation}\label{lagrangian}
{\cal L}=\frac{i}{2}(uu^{\ast}_t-u^{\ast}u_t+vv^{\ast}_t-
v^{\ast}v_t)+\frac{1}{2}|u_x|^2+\frac{1}{2}|v_x|^2+V_1(x)|u|^2+
V_2(x)|v|^2-\frac{g_{11}}{2}|u|^4-\frac{g_{22}}{2}|v|^4 -
g_{12}|u|^2|v|^2,
\end{equation}
yield the effective Lagrangian $L=\int{\cal L}dx$ based on trial
functions (\ref{ansatz}) and harmonic traps
$V_{1,2}=\beta_{1,2}\,x^2$
\begin{eqnarray}
L &=&
\frac{N_1}{4a_1^2}+\frac{N_2}{4a_2^2}-\frac{g_{11}N_1^2}{2\sqrt{2\pi}a_1}
-\frac{g_{22}N_2^2}{2\sqrt{2\pi}a_2}+N_1a_1^2b_1^2+N_2a_2^2b_2^2+
\frac{1}{2}N_1a_1^2b_{1t}+\frac{1}{2}N_2a_2^2b_{2t}-
\\ & & \frac{g_{12}N_1N_2}{\sqrt{\pi}(a_1^2+a_2^2)^{1/2}}+
\frac{1}{2}\beta_1N_1a_1^2+\frac{1}{2}\beta_2N_2a_2^2+N_1\phi_{1t}+N_2\phi_{2t}.
\end{eqnarray}
The equations for the widths are derived from Euler-Lagrange
equations $d/dt(\partial L/\partial \dot{a}_i)-\partial L/\partial
a_i=0$, $i=1,2$.
\begin{eqnarray}
\ddot{a}_1&=&\frac{1}{a_1^3}-\frac{g_{11}
N_1}{\sqrt{2\pi}a_1^2}-\frac{2 g_{12} N_2
a_1}{\sqrt{\pi}(a_1^2+a_2^2)^{3/2}} - 2\, \beta_1 a_1, \label{a1tt}\\
\ddot{a}_2&=&\frac{1}{a_2^3}-\frac{g_{22}
N_2}{\sqrt{2\pi}a_2^2}-\frac{2 g_{12} N_1
a_2}{\sqrt{\pi}(a_1^2+a_2^2)^{3/2}} - 2\, \beta_2 a_2, \label{a2tt}
\end{eqnarray}
where the over-dot implies the time derivative, $N_{1,2}=A_{1,2}^2
a_{1,2} \sqrt{\pi}$ is the norm of the corresponding component. The
fixed point of this system allows us to find the stationary waveform
and explore its small amplitude dynamics.

\begin{figure}[htb]
\centerline{
\includegraphics[width=5cm,height=5cm,clip]{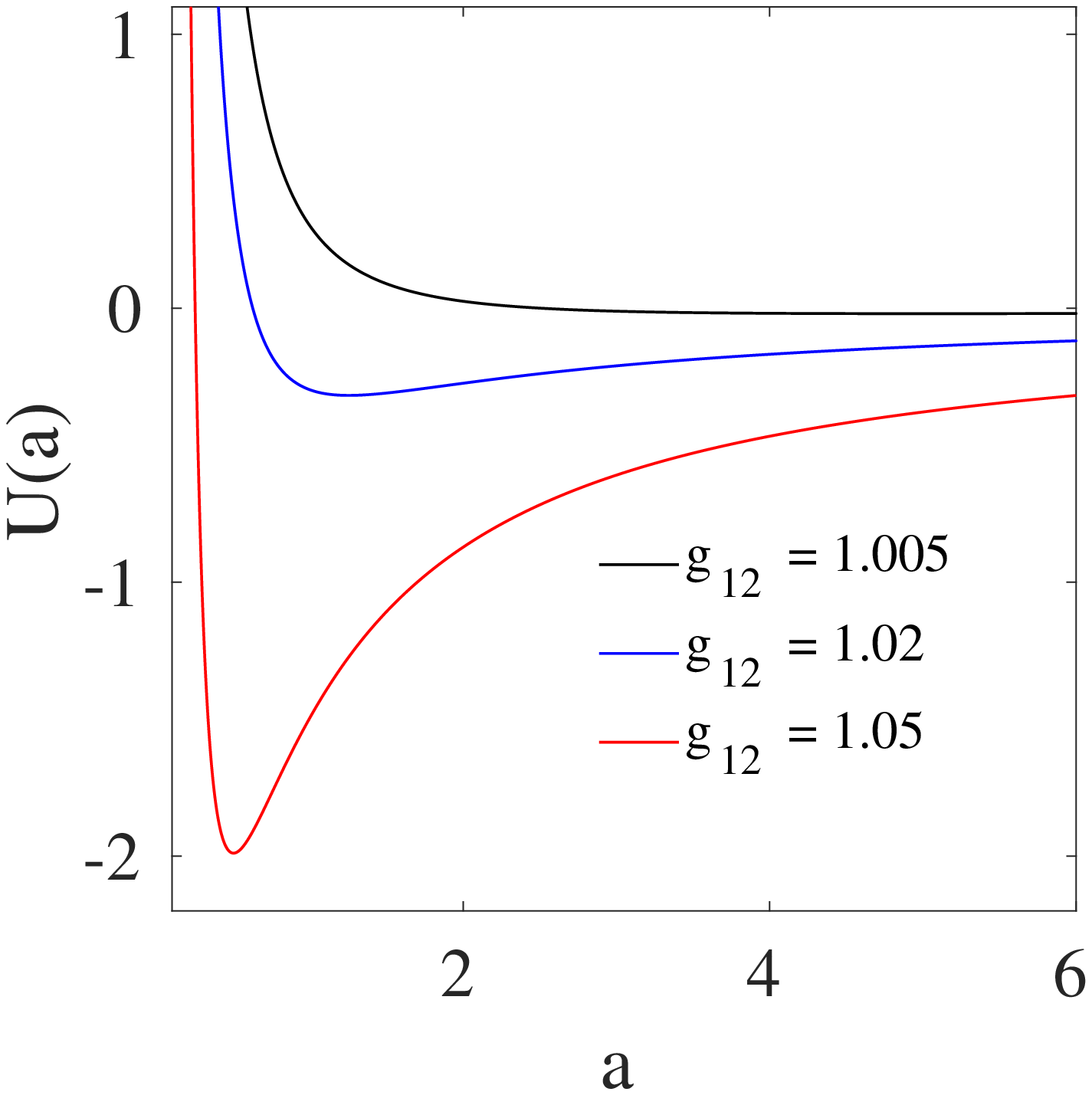}\quad
\includegraphics[width=5cm,height=5cm,clip]{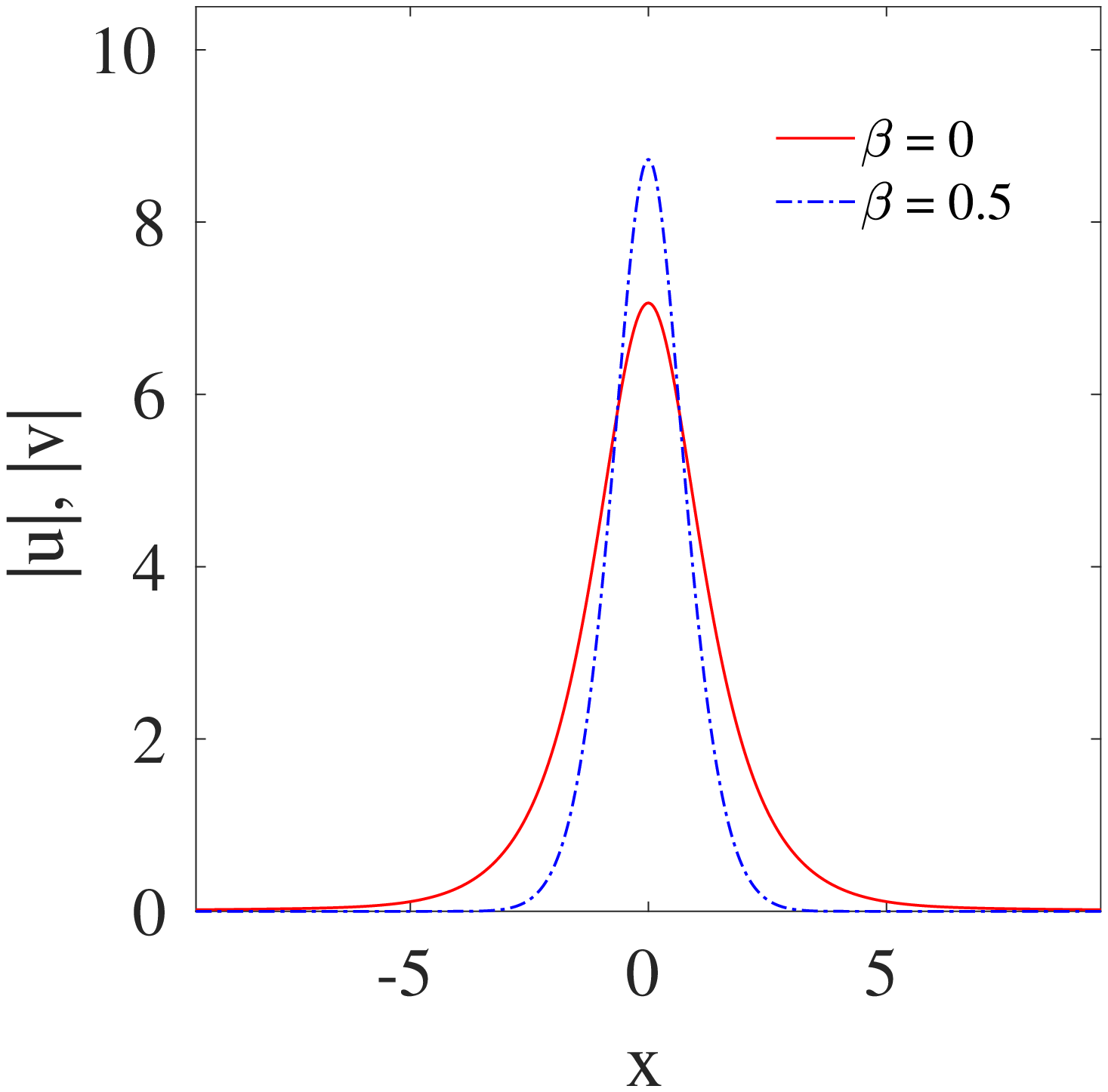}\quad
\includegraphics[width=5cm,height=5cm,clip]{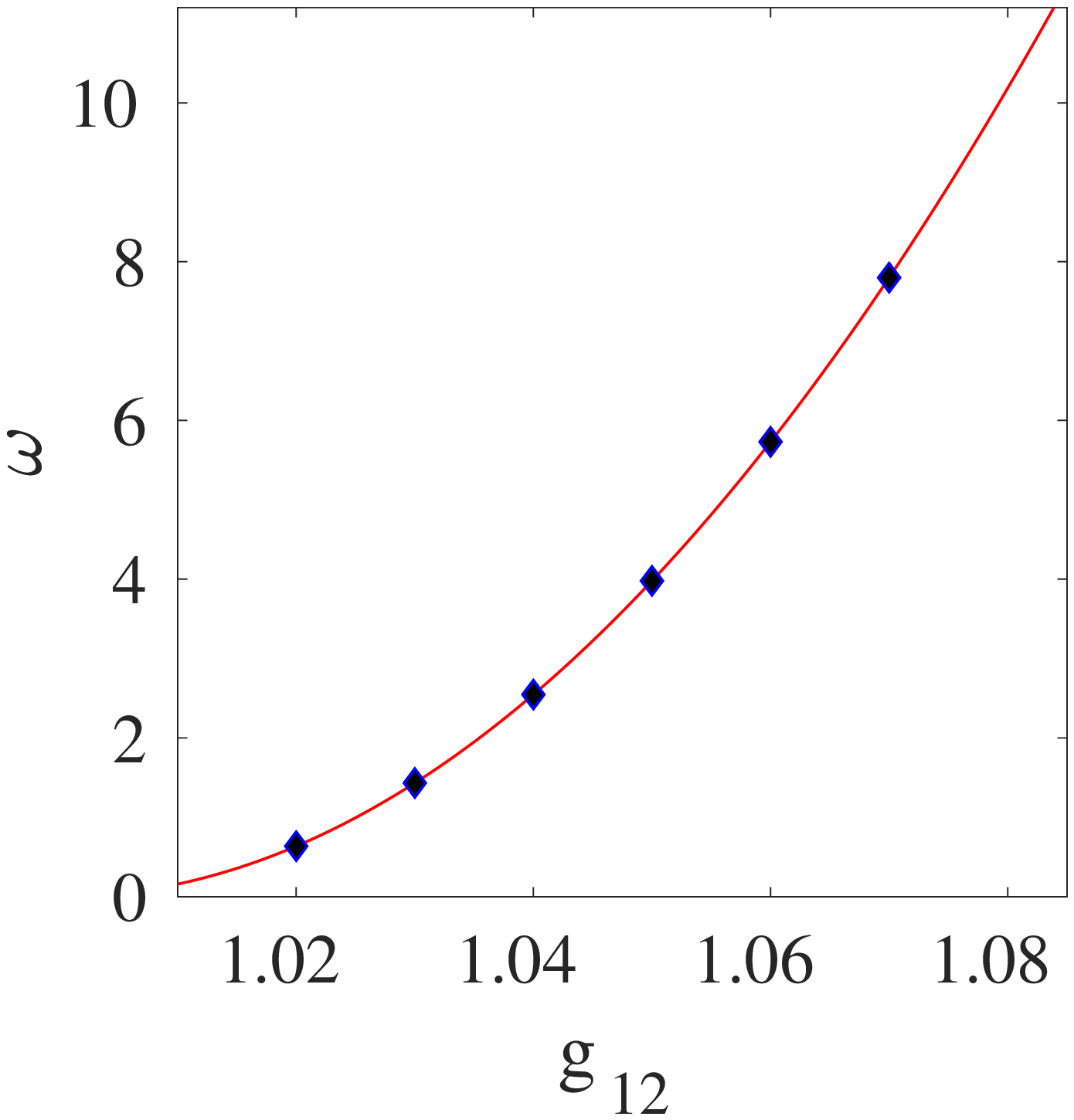}}
\caption{Left panel: A local minimum in the potential $U(a)$
signifies the presence of a bound state of two self-repulsive BECs,
held together only thanks to the inter-species attraction in free
space ($\beta=0$). When the inter-component attraction is
sufficiently weak the local minimum disappears (black solid line),
therefore the bound state doesn't exist. Middle panel: The shapes of
stable symbiotic localized states in absence (red solid line) and in
presence (blue dashed line) of a parabolic potential for
$g_{12}=1.02$. Right panel: The frequency of internal vibrations of
a symbiotic localized state as a function of the inter-component
coupling according to Eq. (\ref{omega}) (solid line) and GPE
numerical results (symbols). Parameter values: $N=N_1=N_2=100$,
$g=g_{11}=g_{22}=-1$.} \label{f1}
\end{figure}

\begin{figure}[htb]
\centerline{
\includegraphics[width=5cm,height=5cm,clip]{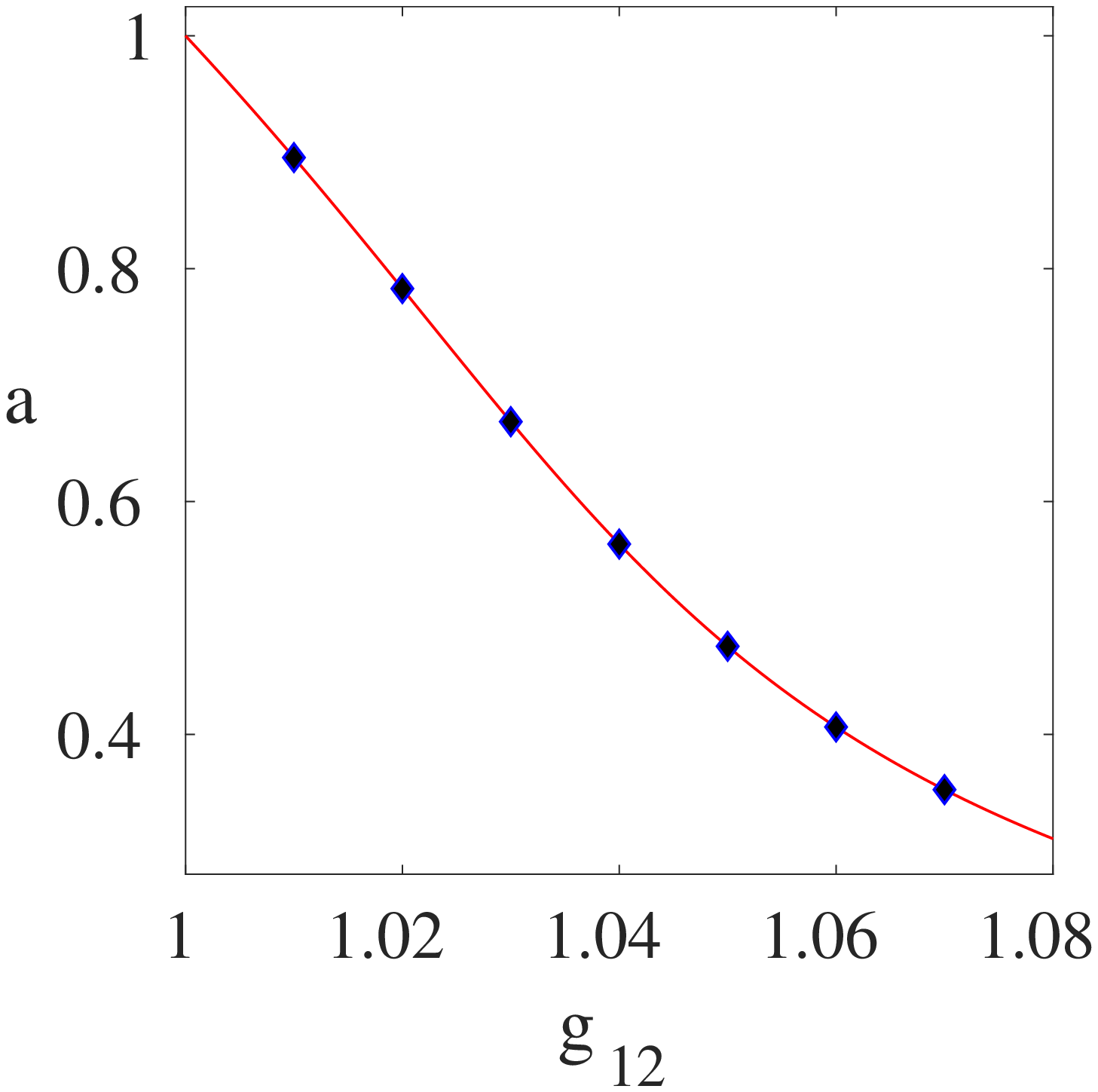}\quad
\includegraphics[width=5cm,height=5cm,clip]{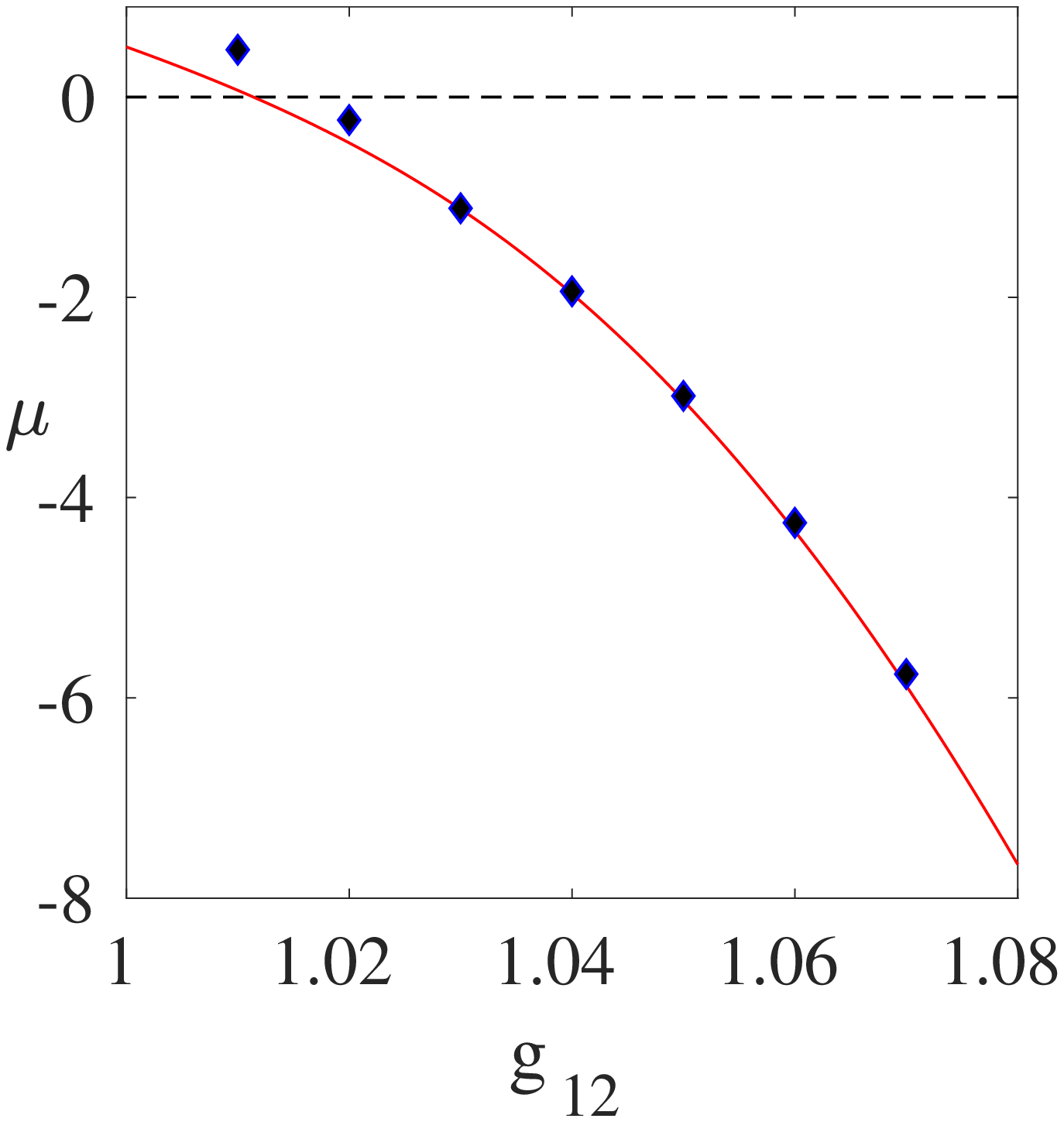}\quad
\includegraphics[width=5cm,height=5cm,clip]{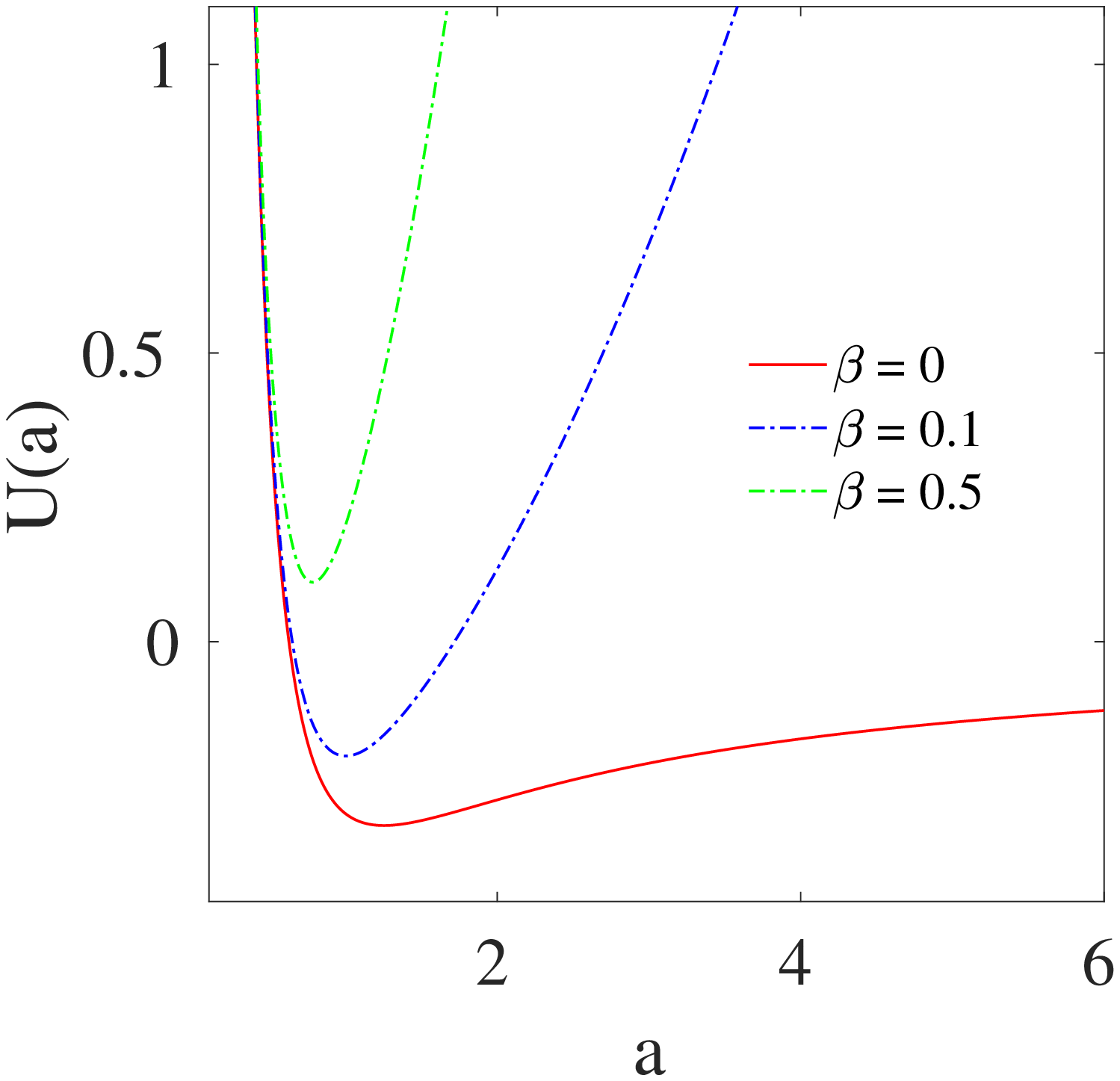}}
\caption{Left panel: The localized state shrinks as the
inter-component coupling becomes stronger according to Eq.
(\ref{aext}). Middle panel: The chemical potential crosses zero
around $g_{12}\simeq 1.01$ signaling the decay of the symbiotic
soliton at weaker inter-component attraction for $\beta=0.5$. This
corresponds to potential Eq. (\ref{pot}) being positive at its
minimum. Right panel: When the harmonic trap is present ($\beta
\not= 0$), the minimum in the potential $U(a)$ always exists.
Negative energy at minimum (blue dashed line) corresponds to a bound
state, while positive energy (green dashed line) is evidence of the
bound state decay. Parameter values are similar to previous figure.
Symbols correspond to numerical GPE simulations. } \label{f2}
\end{figure}

Most convenient for analytical treatment is the above mentioned
symmetric setting $a=a_1=a_2$, $N=N_1=N_2$, $g=g_{11}=g_{22}$ with
similar external potentials $\beta=\beta_1 = \beta_2$. Then the
system Eqs.(\ref{a1tt})-(\ref{a2tt}) reduces to a single equation
\begin{equation}\label{datt}
\ddot{a}=\frac{1}{a^3}-(g+g_{12})\frac{N}{\sqrt{2\pi} a^2}-2\beta a.
\end{equation}
The existence of stable bound state of two self-repulsive BECs
($g<0$) held together solely by inter-component attraction
($g_{12}>0$), even in the absence of trap ($\beta=0$), is evident
from the shape of the potential
\begin{equation}\label{pot}
U(a)=\frac{1}{2a^2}-(g+g_{12})\frac{N}{\sqrt{2 \pi} a} + \beta a^2,
\end{equation}
which corresponds to above equation of motion $\ddot{a}=-\partial
U/\partial a$, as illustrated in Fig. \ref{f1}. In presence of
harmonic potential ($\beta\not= 0$) the width of the stationary
localized state can be found from solution of the algebraic equation
\begin{equation}\label{aext}
a^4+p\,a+q=0, \quad p=\frac{(g+g_{12})N}{2\,\sqrt{2\pi}\beta}, \quad
q=-\frac{1}{2\beta},
\end{equation}
whose real and positive roots are\cite{korn-book}
\begin{equation}\label{aext1}
a=-\frac{1}{2}\left(s^{1/2}-\sqrt{2ps^{-1/2}-s}\right), \quad
s=\frac{4 q}{3 r}, \quad r=\left(\frac{p^2}{2} +
\frac{1}{2}\sqrt{p^4-\frac{256}{27}q^3}\right)^{1/3}.
\end{equation}
The corresponding chemical potential of the symbiotic soliton
confined to a parabolic potential can be found using a static
variational approach
\begin{equation}\label{mu}
\mu=\beta a^2-(g + g_{12})\frac{3 N}{4\sqrt{2\pi} a}.
\end{equation}
The width of the trapped symbiotic soliton and its chemical
potential according to Eqs. (\ref{aext})-(\ref{mu}) are shown in
Fig.~\ref{f2}, which is drawn as follows. First we calculate the
width $a$ from Eq. (\ref{aext1}) for given inter-component coupling
parameter $g_{12}$, then use it in Eq. (\ref{mu}). Symbols in these
figures correspond to numerical data, which are found from GPE as
\begin{equation}\label{agpe}
a_1(t)=\left(2\int_{-\infty}^{\infty} x^2 |u(x,t)|^2 dx /
\int_{-\infty}^{\infty} |u(x,t)|^2 dx \right)^{1/2},
\end{equation}
and similarly for $a_2(t)$. Numerical data for the chemical
potential are found from oscillation frequency of the stationary
solution $Re(u(x,t))$. It should be noted, that although the
symbiotic soliton does'n exist for $g_{12} \leq 1.01$, the localized
state retains its gaussian shape owing to the harmonic trap
confinement of the BEC.

Imbalanced mixtures of binary BECs require an external potential for
the unbound excess atoms of the majority component to remain
confined in space. In fact, the bigger component can act as a
trapping potential for the smaller one. Therefore, trapping only the
bigger component ($V_1 = 0, V_2\not=0$) is sufficient for the
existence of stable symbiotic localized states in an external
potential, as shown in Fig. \ref{f2}.

\section{Numerical results}

Small amplitude vibrations of a free ($\beta=0$) symbiotic soliton
near the equilibrium $a_0=\sqrt{2\pi}/[(g+g_{12}) N]$ can be
investigated by representing $a(t)=a_0+\delta a(t)$ with $\delta a
\ll a_0$. Expansion of the right-hand side of Eq.(\ref{datt}) and
keeping the terms up to the first order in $\delta a$ yields the
harmonic oscillator equation $\delta a_{tt}+\omega^2 \delta a = 0$
with corresponding frequency:
\begin{equation}\label{omega}
\omega =(g+g_{12})^2 N^2/(2\pi).
\end{equation}
The Eq.(\ref{omega}) connects the frequency of the soliton
vibrations, $\omega$, with the coefficient of the inter-component
coupling $g_{12}$. This relation is verified by direct comparison
with GPE numerical simulations, as shown in Fig. \ref{f1}c. To
induce vibrations on the soliton profiles a slight change (by one
percent) of the coefficient $g_{12}$ has been employed. At stronger
inter-component attraction ($g_{12} \simeq 1.1$) the oscillations
become highly anharmonic, that is why we show the GPE results up to
$g_{12}=1.08$.

We remark that in the literature the properties of symbiotic
localized states are considered mainly for balanced settings, when
the two components are self-trapped in free space or stabilized by a
periodic potential
\cite{adhikari2005,perezgarcia2005,adhikari2008,javed2022}. The
imbalanced case, when the number of atoms in one component is
significantly greater than that in the other component, remains less
explored. In this connection, the motion of a symbiotic localized
state in the surrounding gas of unbound atoms of the majority
component represents particular interest.

The analytic approach for strongly imbalanced binary mixtures is
complicated by the presence of the external potential, which is
necessary to prevent the spreading of the unbound part of the
majority component, and by unusual waveforms of resulting symbiotic
states, this making the choice of a suitable ansatz for the
variational method more difficult. Below we produce symbiotic
localized states in trapped binary self-repulsive BECs with
attractive inter-component interaction by numerical methods. In
particular, external potentials in the form of a soft-wall box
\begin{equation}\label{box}
V(x)=\frac{V_0}{2}\left[{\rm th}\left(\frac{x-h}{w}\right) - {\rm
th}\left(\frac{x+h}{w}\right)+2\right],
\end{equation}
where the parameters $V_0, h, w$ characterize its strength, spatial
extent and width of the transition region, and in the form of a
parabolic trap $V(x)=\beta \, x^2$, will be considered. In both
cases the external potential is necessary only for the majority
component, to prevent its excess atoms from spreading. Note that the
minority component, being immersed fully in the majority one,
doesn't require any external potential. All of its atoms are bound
due to inter-species attraction, i.e. the majority component acts as
a confining medium for the minority one. A similar situation was
shown to occur also in the context of a BEC quantum dot
\cite{salerno2005}.

\begin{figure}[htb]
\centerline{\qquad a) \hspace{6cm} b)} \centerline{
\includegraphics[width=6cm,height=5cm,clip]{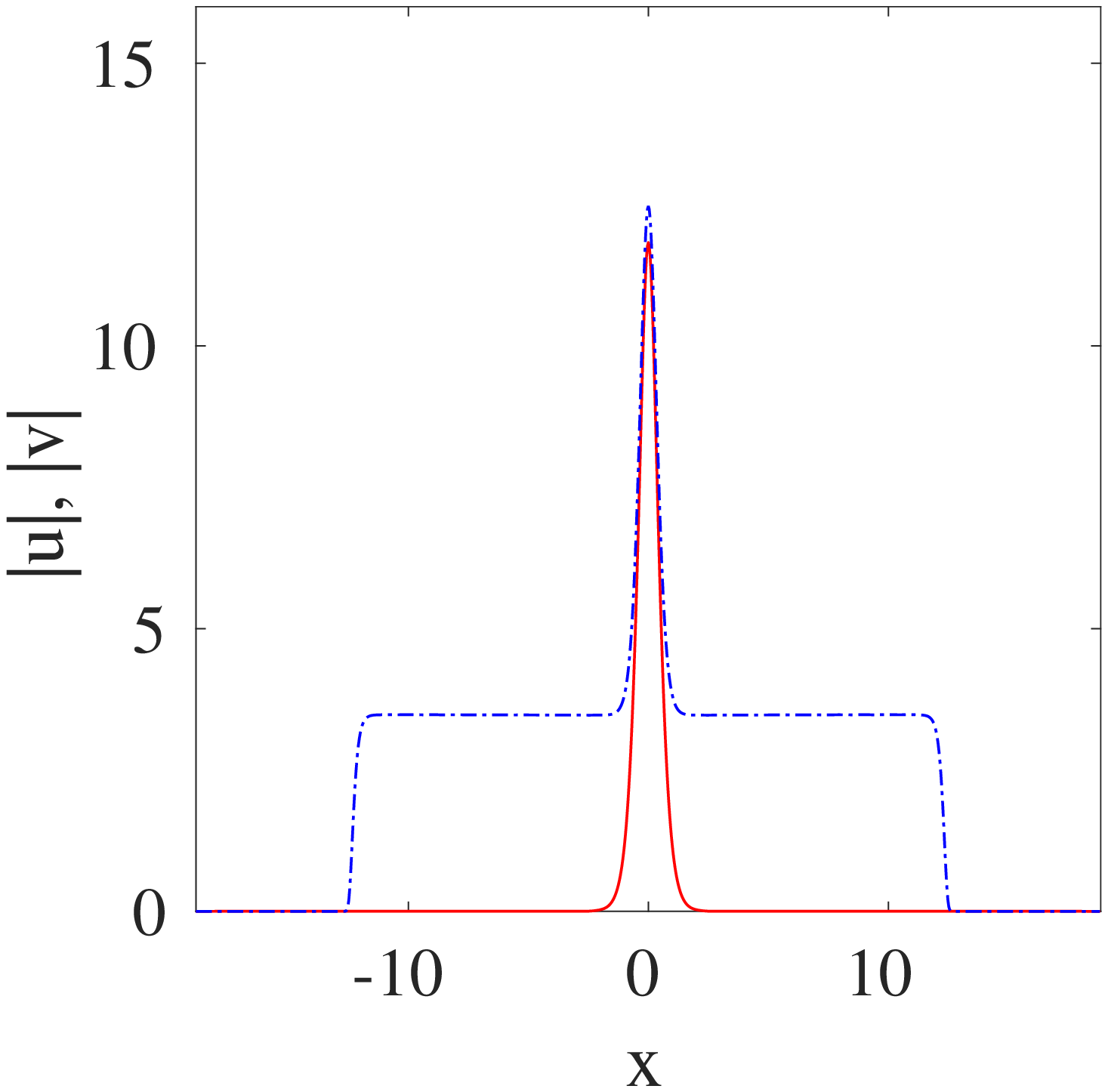}\qquad
\includegraphics[width=6cm,height=5cm,clip]{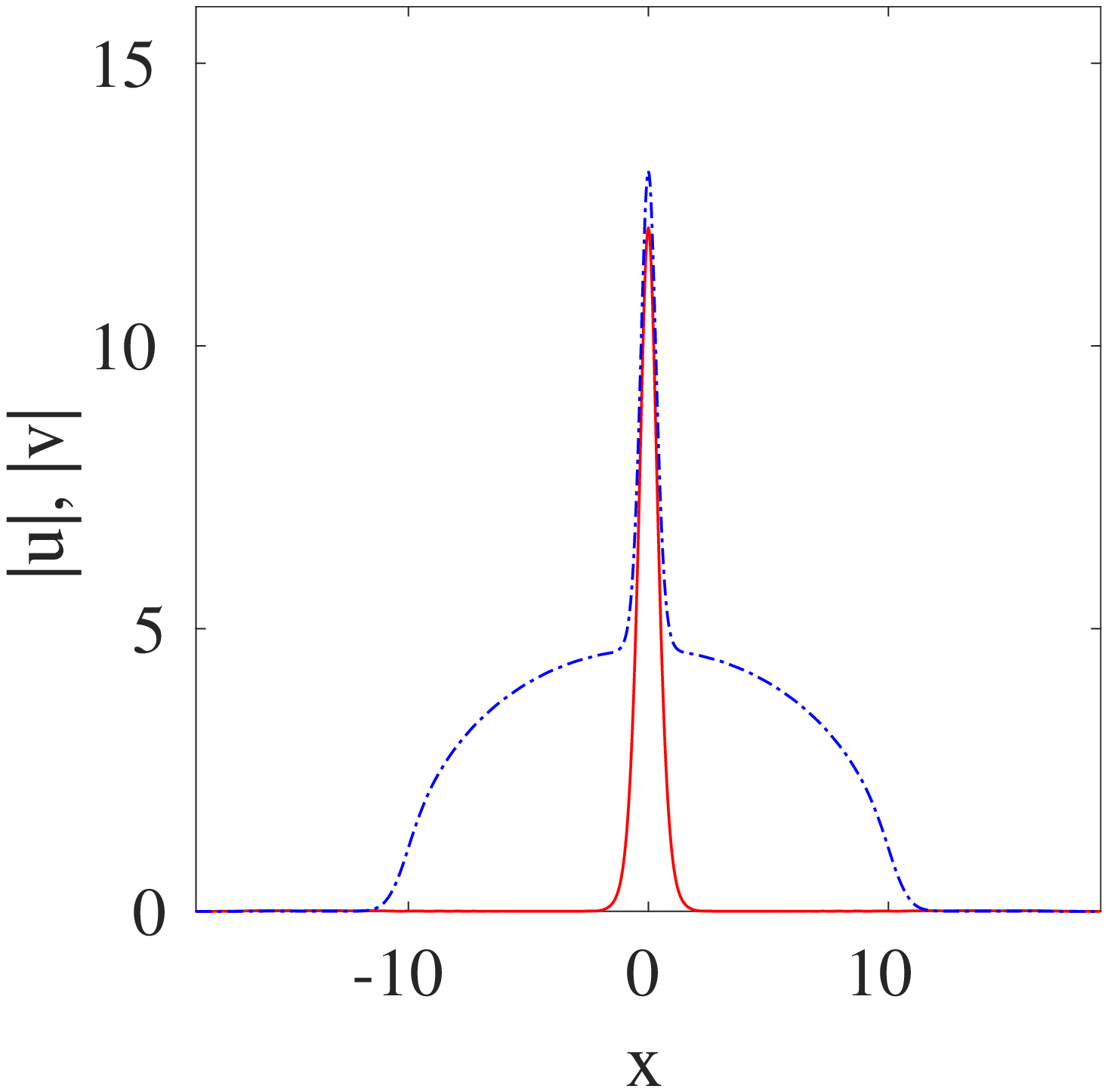}}
\vspace{0.1cm} \centerline{\qquad c) \hspace{6cm}d)} \centerline{
\includegraphics[width=6cm,height=5cm,clip]{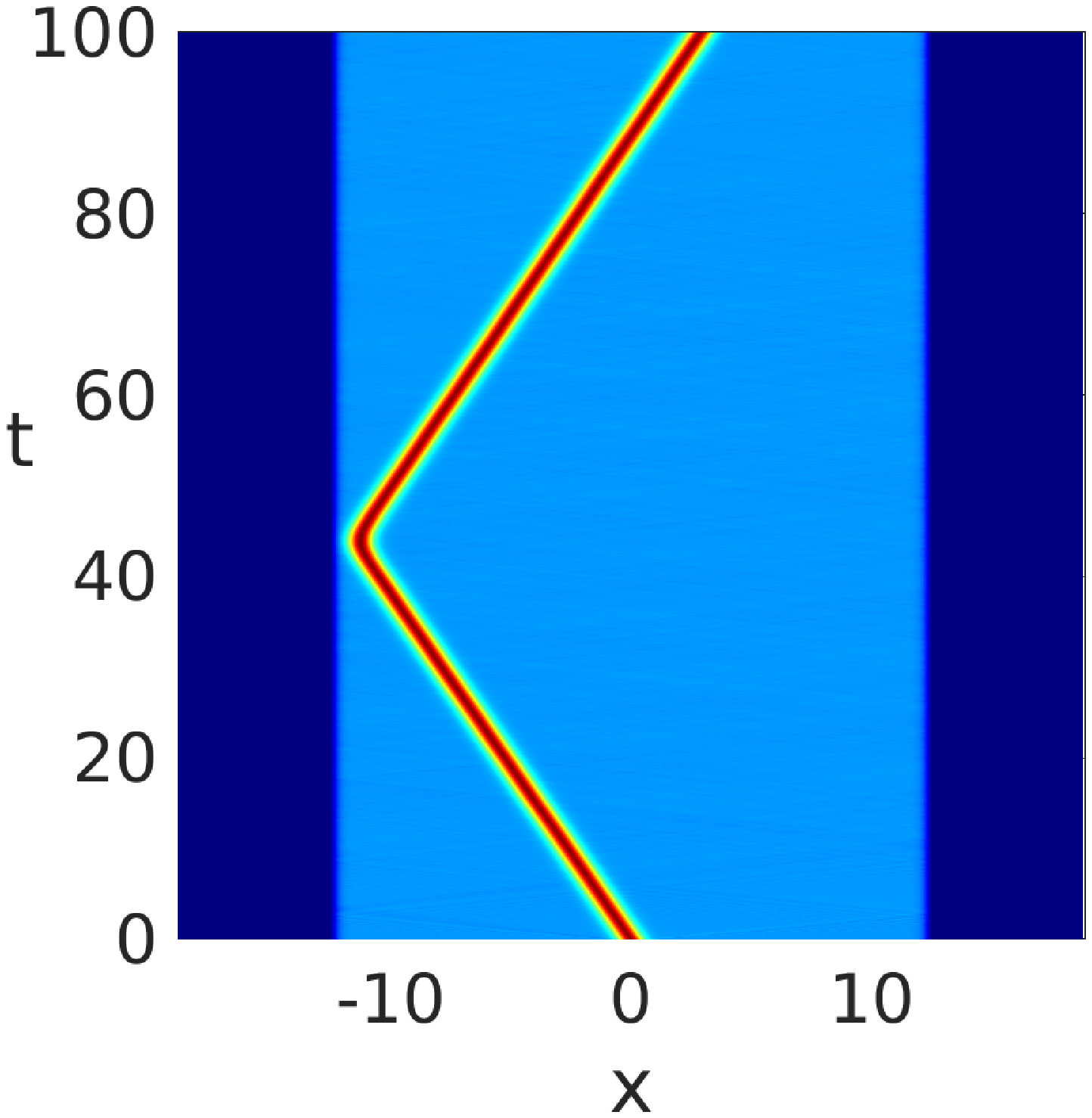}\qquad
\includegraphics[width=6cm,height=5cm,clip]{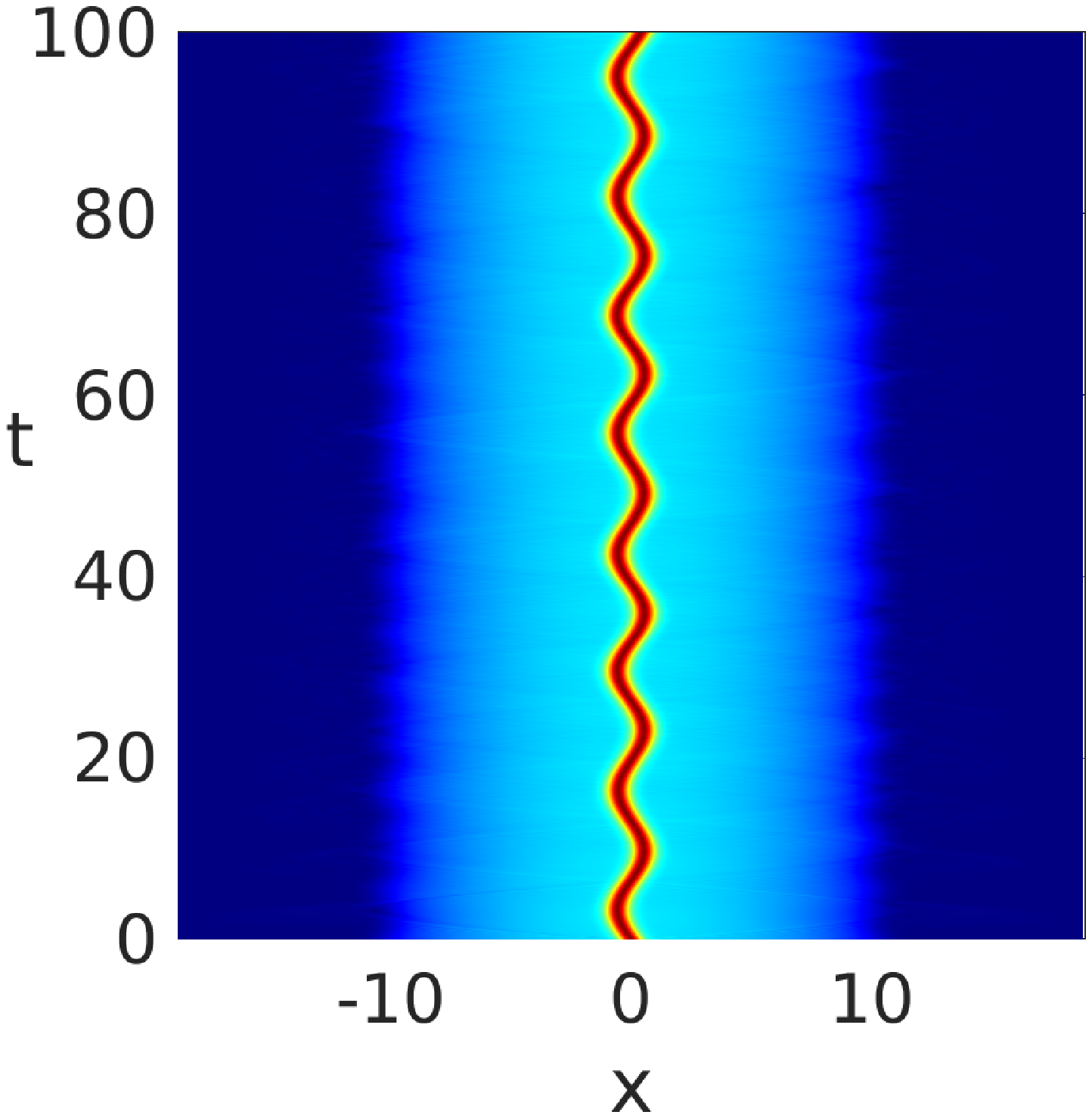}}
\caption{Upper panels: Stable symbiotic localized states in binary
self-repulsive BECs with inter-species attraction, trapped in a
box-like potential (a) and harmonic trap (b). Lower panels: Density
plots showing the motion of the symbiotic localized state as a whole
through the bath of unbound atomic gas of the majority component
(light blue region). In the box-potential the symbiotic soliton
moves along the straight line with constant velocity until it
reflects from the boundary (c). In the case of parabolic trap, the
motion is oscillatory with a constant period (d). The minority
component is set in motion with velocity $v=0.5$. Parameter values:
$g_{11}=g_{22}=-1$, $g_{12}=1.05$, $N_1=100$, $N_2=400$, $V_0=1000$,
$h=4\pi$, $w=0.1$, $\beta_1=0$, $\beta_2=0.2$.} \label{f3}
\end{figure}

The waveforms of typical symbiotic localized states with imbalanced
atom numbers in the two components ($N_2 = 4 N_1$), and their
dynamics are shown Fig. \ref{f3}. As it can be seen from the bottom
panels of this figure, in both cases the background medium (unbound
part of the majority component) remains unperturbed by a moving
soliton with velocity below the Landau critical value as expected
for a superfluid (see below for the case of $v$ above the Landau
critical value). To induce the dynamics of the symbiotic localized
state it is sufficient to set in motion only the minority component
by means of phase imprinting $u \rightarrow u\exp{(ivx)}$.

\begin{figure}[htb]
\centerline{
\includegraphics[width=6cm,height=6cm,clip]{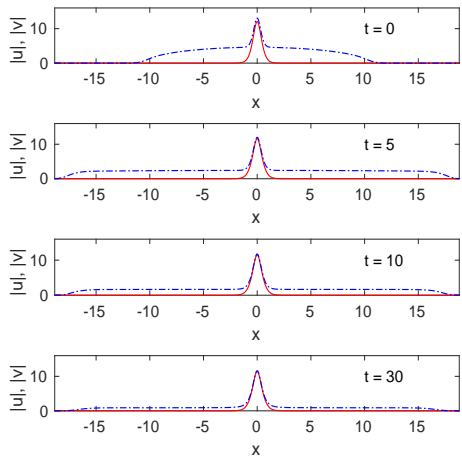}}
\caption{Evolution of the imbalanced self-repulsive binary BEC when
the harmonic trap is removed at $t=0$. Excess atoms of the majority
component (blue dashed line) move away from the center and get
absorbed at domain boundaries. A pure symbiotic soliton emerges at
$t=30$. } \label{f4}
\end{figure}

Note that in our model, the symbiotic soliton is embedded in the
background gas of unbound atoms of the bigger component. When the
trapping potential is removed, the excess atoms of the imbalanced
mixture move away from the center, subsequently being absorbed by
the domain boundaries, and pure symbiotic soliton emerges, as
illustrated in Fig. \ref{f4}.

\subsection{All-repulsive interactions}

\begin{figure}[htb]
\centerline{
\includegraphics[width=6cm,height=5cm,clip]{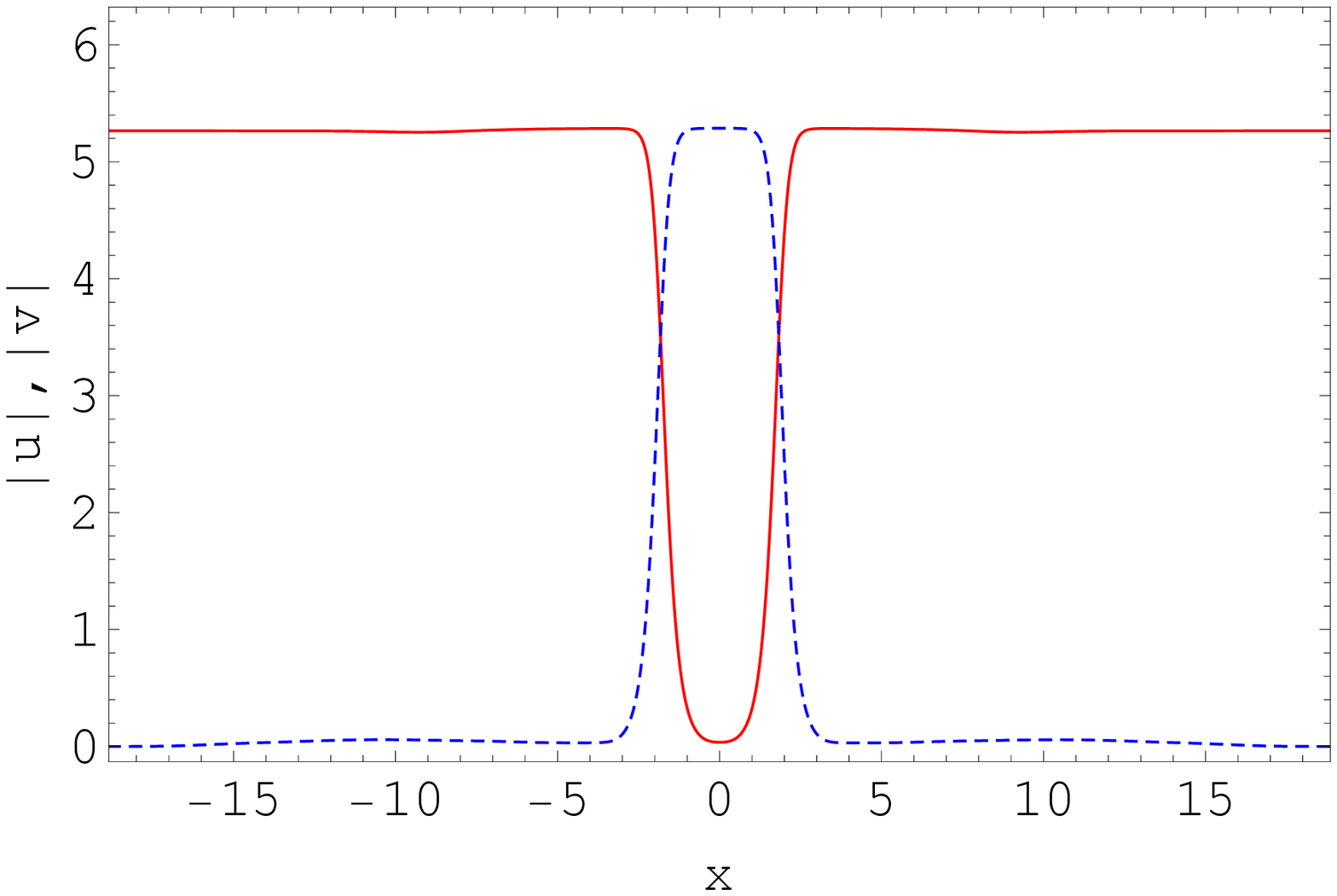}\qquad
\includegraphics[width=6cm,height=5cm,clip]{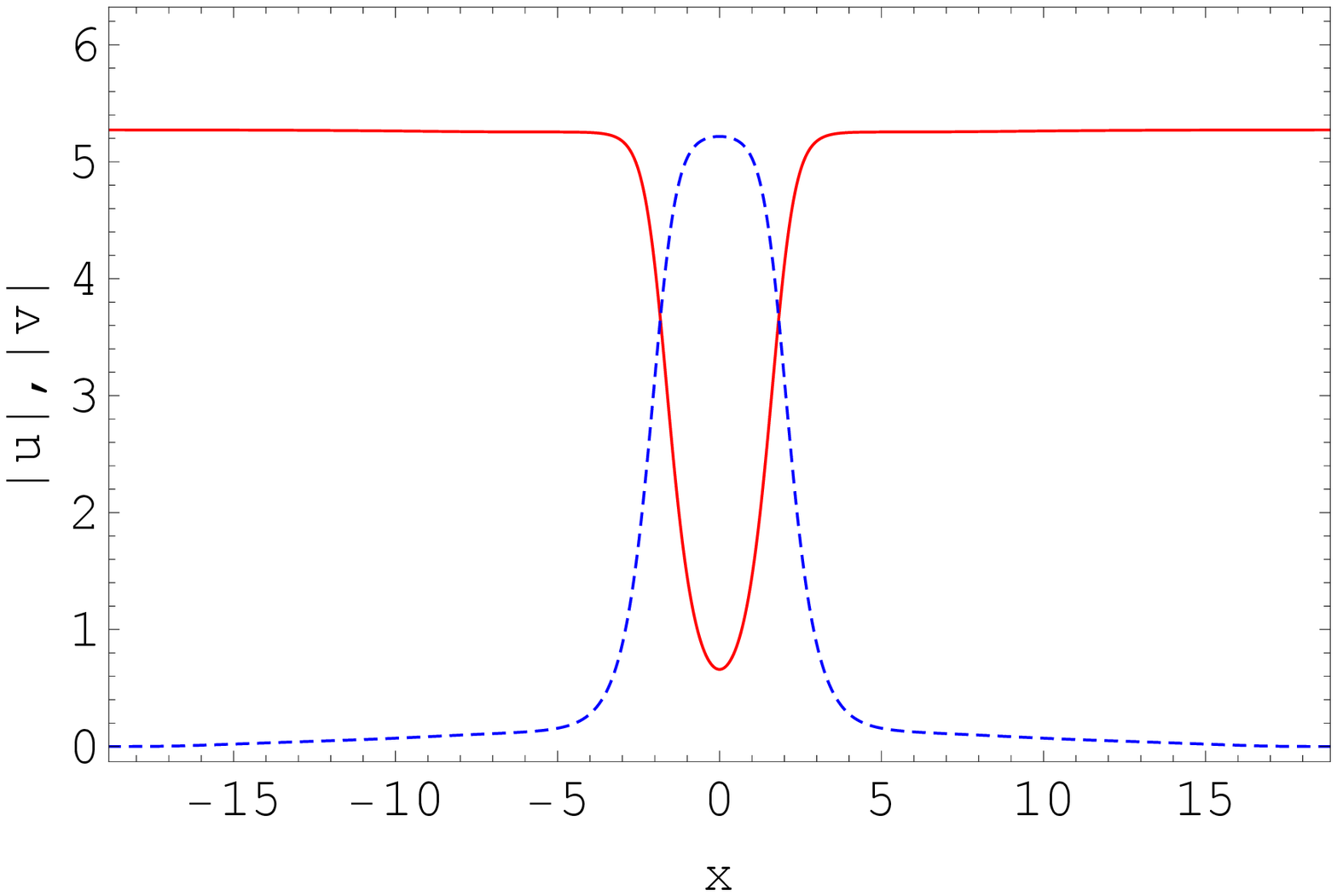}}
\caption{Symbiotic soliton profiles of a binary condensate with
all-repulsive interactions confined to a ring-shaped quasi-1D trap.
Numerical simulations of the GPE (\ref{u})-(\ref{v}) using the
Pitaevskii damping procedure \cite{choi1998} are performed with
periodic boundary conditions. Starting wave profiles correspond to a
flat background for the majority component with amplitude $u(x,0)=5$
and a Gaussian function for the minority component $v(x,0)=A
\exp(-x^2/(2 a^2))$ with $A = 6.71, a = 1.25$. In the course of
evolution with a phenomenological damping term in the GPE a stable
symbiotic soliton emerges. To improve numerical convergence the
absorbing boundary conditions has been employed for the localized
component $v(x,t)$ only. Other parameters are fixed as:
$g_{11}=g_{22}=-1$, $g_{12}=-1.05$ (left panel) and $g_{12}=-1.02$
(right panel). } \label{f5}
\end{figure}

Now we consider a special case of \textit{all-repulsive}
interactions, when the mixture of two self-repulsive condensates
($g_{11}<0, g_{22}<0$) with inter-species repulsion ($g_{12}<0$) is
confined to a ring-type quasi-1D potential. In numerical simulations
such a setting is modeled via periodic boundary conditions
$u(0,t)=u(L,t), \ v(0,t)=v(L,t)$, with $L$ being the length of the
integration domain. It should be noted that ground states of binary
condensates with all-repulsive atomic interactions confined to 1D
box potentials was reported in Ref. \cite{parajuli2019}. In our
setting the density distortions caused by hard walls of the box
potential \cite{parajuli2019} have been avoided and the ring
geometry enables continuous circulation of the symbiotic soliton. In
Fig. \ref{f5} the ground state wave profiles of a two-component BEC
with all-repulsive interactions is illustrated. As it can be seen,
the stronger cross-repulsion leads to a full depletion of the
background component at the origin, while a weaker cross-repulsion
allows the non-zero amplitude of both components within the overlap
region. It should be emphasized that although these localized modes
resemble the well known dark-bright soliton, they cannot exist in
isolation and therefore belong to the class of symbiotic solitons.
Unlike previously reported symbiotic solitons existing due to
inter-component attraction, these localized modes exist only due to
inter-component repulsion. Similar solutions were also discussed in
\cite{filatrella2014} for the cases of immiscible binary BEC
mixtures and Tonks-Girardeau gases. When a stable symbiotic soliton
in the all repulsive setting has been created, the dynamics can be
induced by setting in motion the minority component with some
velocity.

\begin{figure}[htb]
\centerline{
\includegraphics[width=6cm,height=5cm,clip]{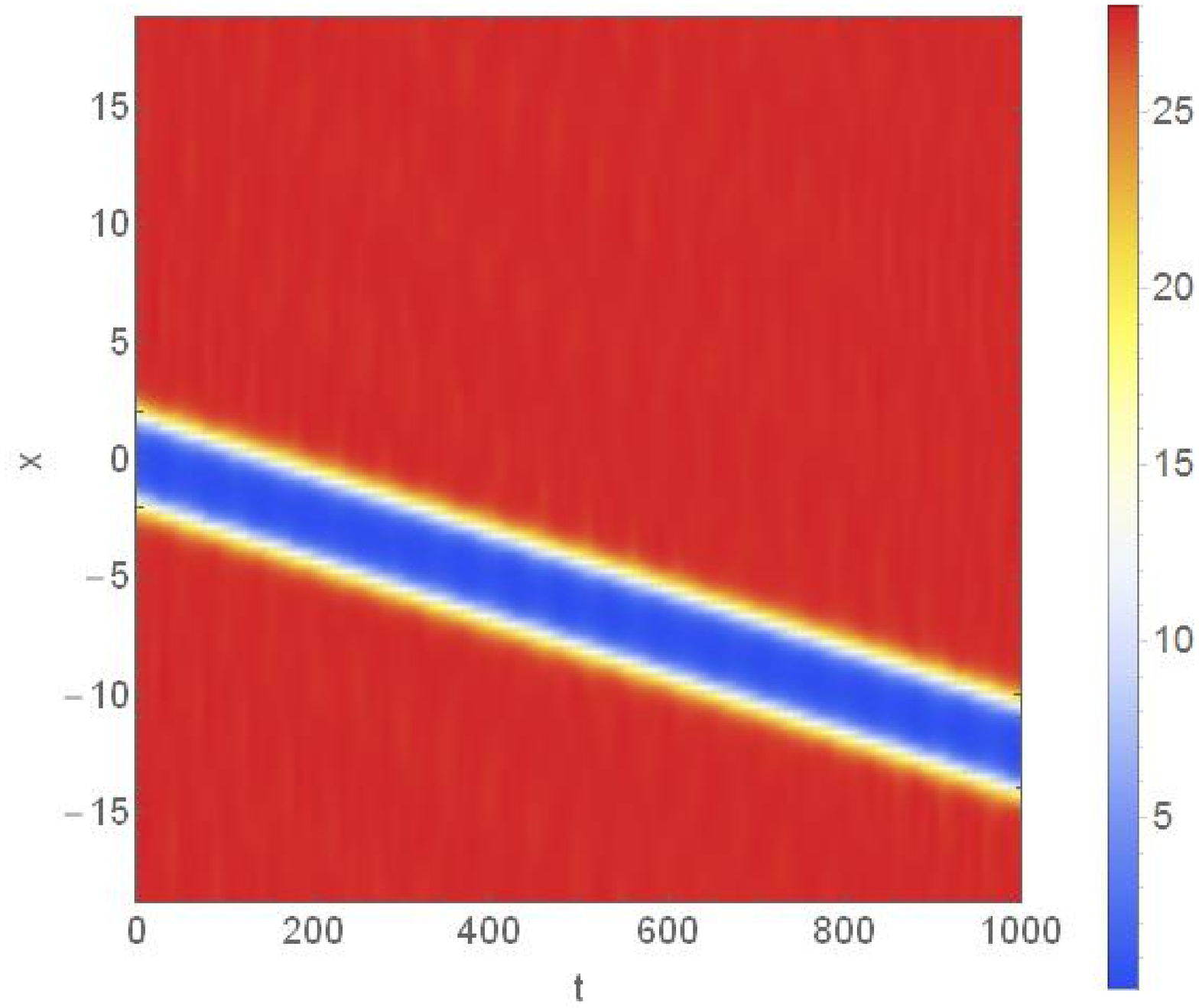}\qquad
\includegraphics[width=6cm,height=5cm,clip]{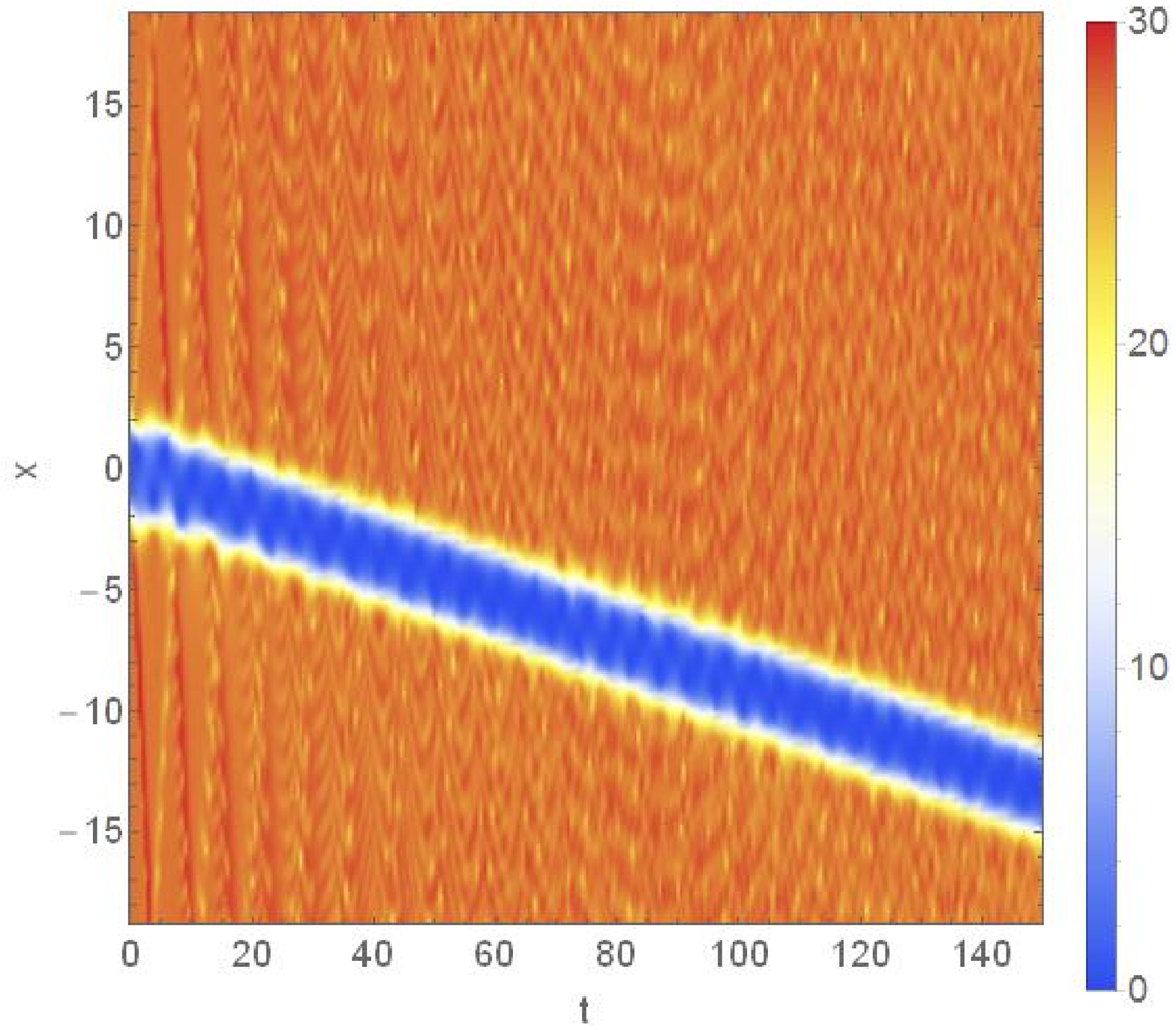}}
\caption{Density plots for the majority component $|u(x,t)|^2$
according to the numerical solution of the GPE (\ref{u})-(\ref{v}).
At sub-critical velocity ($v=0.1$) the background condensate remains
uniform (left panel) which is evidence of its superfluid property.
In contrast, at super-critical velocity ($v=1$) notable density
modulations emerge (right panel), signaling the disruption of the
superfluidity. Parameter values correspond to strength of
cross-repulsion ($g_{12}=-1.05$) slightly exceeding that of the
self-repulsion ($g_{11}=g_{22}=-1$). } \label{f6}
\end{figure}

To illustrate superfluid property of the system we set in motion the
minority component with different velocities and observe emerging
density modulations on top of the majority component. The
characteristic example is shown in Fig. \ref{f6}. When the velocity
is smaller than the speed of sound for the background condensate
$v<v_s=\sqrt{g_{3D} n/m}$, the excitations have not been produced,
while for the velocity exceeding the speed of sound ($v>v_s$)
density perturbations show up. These observations are in line with
the Landau theory of superfluidity \cite{pitaevskii-book}.

\subsection{Relevance to experimental conditions}

The experimental realization of a binary BEC in a toroidal trap
\cite{beattie2013} opened new possibilities in exploring the
superfluidity in multi-component systems. The condensate was created
using $^{87}$Rb atoms in two different spin states $|F=1,
m_F=1\rangle$ and $|F=1, m_F=0\rangle$, each component containing $N
\sim 10^4$ atoms. The ring trap with transverse frequency
$\omega_{\bot}=2 \pi \times 50$ Hz allowed tight confinement of the
condensate so that the dynamics along the ring can be considered in
quasi-1D approximation. If the average radius of the ring is $r
\simeq 12 \ \mu$m \cite{chen2019} the length of the circular trap
will be ${\cal L}=2 \pi r \simeq 75 \ \mu$m. The same-species and
cross-species $s$-wave scattering lengths for $^{87}$Rb slightly
differ from each other \cite{egorov2013} and can be estimated in
units of Bohr radius ($a_B$) as $a_{11}=a_{22} \simeq -100 \times
a_B$ and $a_{12} \simeq -98 \times a_B$, these parameters being
tunable by the Feshbach resonance method \cite{chin2010}.

Now using the atomic mass of $^{87}$Ru $m=1.4\times 10^{-25}$ kg and
the transverse frequency $\omega_{\bot}=2 \pi \times 50$ Hz $=$ 314
sec$^{-1}$, we can estimate the unit of length employed in numerical
simulations $l_{\bot}=\sqrt{\hbar/m \omega_{\bot}} \simeq 1.5 \
\mu$m. The integration domain length corresponds to $L=12 \pi
l_{\bot}\simeq 58 \ \mu$m, while the volume occupied by the
condensate is $V=L l_{\bot}^2 \simeq 1.4 \times 10^{-16}$m$^{-3}$.
The atomic density is estimated as $n=N/V \simeq 7.3 \times
10^{19}$m$^{-3}$. The coupling strength for the majority component
$g_{3D}=4\pi \hbar^2 |a_{11}|/m \simeq 5.2 \times 10^{-51}$ kg
m$^5$/sec$^2$. Using the above parameters we can estimate the speed
of sound $v_s = \sqrt{g_{3D} n /m} \simeq 1.6 \times 10^{-3}$m/sec,
which in dimensionless units corresponds to $v_s/(\omega_{\bot}
l_{\bot}) \simeq 3$. The above presented estimates show, that the
parameter values used in our numerical simulations qualitatively
correspond to experimental conditions.

\subsection{Details of numerical simulation}

Numerical solution of the governing Eqs. (\ref{u})-(\ref{v}) has
been performed using the method of fast Fourier
transform~\cite{agrawal-book} with 1024 Fourier modes
\cite{numrecipes} within the integration domain $x \in [-6\pi,6\pi]$
and with a time step $\delta t=0.0005$. To find the ground state of
imbalanced symbiotic solitons in external potentials we employ the
Pitaevskii phenomenological damping procedure \cite{choi1998} and
Nijhof's method \cite{nijhof2000}, adapted for two-component GPE. We
employed absorbing boundary method to emulate the condition of
infinite integration domain. The accuracy of the numerical
procedures was controlled through conserved quantities, like the
norm, energy, and momentum.

\section{Conclusions}
We have considered static and dynamic properties of symbiotic
localized states in imbalanced binary self-repulsive BECs, confined
to quasi-one-dimensional external potentials. These states can exist
due to inter-component attraction. The slow motion of the symbiotic
localized state in the background gas of the majority component
occurs without exciting perturbations, which is evidence of its
superfluid nature. When the confining potential is removed, the
imbalanced self-trapped localized state smoothly transforms into
pure symbiotic soliton. In the course of evolution, the excess atoms
of the majority component move away from the origin and get absorbed
at the domain boundaries. For binary condensates with all-repulsive
atomic interactions in a ring-type potential we also have found
symbiotic solitons which can exist due to inter-component repulsion.
The developed variational approach allowed us to reveal the
frequency of soliton's vibrations as a function of the intra- and
inter-component coupling parameters. The analytic predictions are
corroborated by numerical GPE simulations.

\section*{Acknowledgements}
KKI acknowledges financial support from the Ministry of Innovative
Development of the Republic of Uzbekistan for a three months grant
under the program "Short Term Scientific Internship of Young
Scientists in Foreign Scientific Organizations", contract No. 74
with the University of Salerno. KKI also acknowledges the Physics
Department "E. R. Caianiello" for the hospitality received and for
the internship opportunity during which this work was completed.

\end{document}